\documentclass[10pt,english,twocolumn,aps,pra]{revtex4}
\usepackage[T1]{fontenc}
\usepackage[latin9]{inputenc}
\usepackage{amsmath}
\usepackage{graphicx}

\makeatletter

\usepackage{float}

\usepackage{amscd}
\usepackage{color} 
\usepackage{bm}

\usepackage{babel}
\makeatother

\begin{document}

\title{Monogamy Inequality and Residual Entanglement of Three Qubits under
Decoherence}

\author{Thiago R. de Oliveira}

\affiliation{Instituto de Física {}``Gleb Wataghin'', Universidade Estadual
de Campinas, 13083-970, Campinas-SP, Brazil}

\begin{abstract}
Exploring an analytical expression for the convex roof of the pure
state squared concurrence for rank 2 mixed states the entanglement
of a system of three particles under decoherence is studied, using
the monogamy inequality for mixed states and the residual entanglement
obtained from it. The monogamy inequality is investigated both for
the concurrence and the negativity in the case of local independent
phase damping channel acting on generalized GHZ states of three particles
and the local independent amplitude damping channel acting on generalized
W state of three particles. It is shown that the bipartite entanglement
between one qubit and the rest has a qualitative similar behavior
to the entanglement between individual qubits, and that the residual
entanglement in terms of the negativity cannot be a good entanglement
measure for mixed states, since it can increase under local decoherence. 
\end{abstract}
\maketitle

\section{Introduction}

The behavior of entanglement in systems of many particles and the
effects of decoherence on it is an important topic and far from being
totally understood. If for one reason this is an issue in studying
the transition from the microscopic to the classical world, it also
has its practical implications since the advent of Quantum Information
and Computation; a scalable quantum computer is likely to need entanglement
at a many particle level.

A major obstacle in these studies is the concept of entanglement per
si, as there is no established theory for multipartite entanglement
(ME) and, even worse, no universal ME quantifier. One possibility
is to analyse the entanglement between all possible bipartitions in
the system (see \cite{Rigolin2006}, for example). But even in this
case there is no easy path, given that most of bipartite entanglement
measures do not have an analytical expression for \emph{mixed} states
of systems with even small Hilbert space dimension. The Entanglement
of Formation and its related quantity, the Concurrence, for example,
are only exactly computable for $2\times2$ systems (two qubits) \citep{Wootters1998}.
An exception is the negativity, which can be obtained in any dimension
\citep{Vidal2002}. However for Hilbert space dimension higher than
$2\times3$ it can underestimate the entanglement: it can be null
when the state is entangled, defining what is called PPT entanglement
or bound entanglement (two recent reviews on entanglement measures
are \citep{Plenio2007,Horodecki2007}
).

Somehow related to this approach to multipartite entanglement and
relevant to it is the property of monogamy: if two qubits are maximally
entangled, neither of them can be entangled, in any way, with a third
one (not even classically correlated$^{\text{1}}$\footnotetext[1]{By the other side a maximmaly classical correlation between A and B, also prohibits any of them of being entangled with a third particle. In PRA 69, 022309 the monogamy between classical and quantum correlation is investigated.}).
But it is in the case where the two particles are only partially entangled
that the monogamy property may be more relevant to multipartite entanglement.
In particular, in the case of three qubits in a pure state Coffman,
Kundu and Wootters \citep{Coffman2000} obtained that $C_{A(BC)}^{2}\geq C_{AB}^{2}+C_{AC}^{2}$,
with $C_{AB}$ ($C_{AC}$) being the concurrence of $A$ with $B$($C$),
while $C_{A(BC)}$ is the concurrence between A and BC, with the last
treated as a single system. This inequality limits A's entanglement
with B and C taken individually and do not allow both of them to increase
indiscriminately, since all the quantities in the inequality varies
from 0 to 1. Besides establishing this inequality Coffman, Kundu and
Wootters also proposed to take the difference, $C_{A(BC)}^{2}-C_{AB}^{2}-C_{AC}^{2}$,
as a tripartite entanglement measure, which they called \emph{residual
entanglement}, $\tau_{ABC}^{c}$. This monogamy inequality is also
valid for the negativity and an analogous tripartite entanglement
measure can be defined \citep{Ou2007}.

Even though for three particles the residual entanglement is well
accepted as a tripartite entanglement measure, most of the studies
of the decoherence of multipartite entanglement only focused on the
bipartite part of this entanglement \citep{Simon2002,Dur2004,Bandyopadhyay2005,Ann2007,Montakhab2008,L'opez2008,Aolita2008}
(see \citep{Carvalho2004} and \citep{Guhne2008} for an exception
and different approach). That happens because in the case of mixed
states despite the fact that the inequality and the residual entanglement
can be established they involve a convex roof optimization of $C_{A(BC)}^{2}$
with no easy general solution.

Here I make use of an analytical expression for the convex roof of
the pure state $C_{A(BC)}^{2}$ for rank two mixed states, obtained
by Osborne \citep{Osborne2005}, to study the monogamy inequality
and the residual entanglement under decoherence for the first time.
Both are investigated in terms of the concurrence and the negativity
for generalized GHZ and W states of three qubits. I also compare the
behavior of the $N_{A(BC)}^{2}$ and its pure state convex roof extension.

\section{Monogamy and Multipartite Entanglement}

One of the first investigations of the monogamy inequality and its
relation to multipartite entanglement was the work of Coffman, Kundu
and Wootters in 2000 \citep{Coffman2000}. There they first showed
that for a \emph{pure state} of three qubits \begin{equation}
4\det\rho_{A}\geq C_{AB}^{2}+C_{AC}^{2},\end{equation}
 with $\rho_{A}$ the reduced density matrix of the qubit $A$. To
get the final inequality in terms of entanglement it is necessary
to make use of two facts: 

\begin{enumerate}
\item the concurrence of a pure state of two qubits, A and B, is given by
$2\sqrt{\det\rho_{A}}$. 
\item even though the state space of AB is four dimension only two of these
dimensions are needed to express a pure state of ABC (this comes from
the Schmidt decomposition of the state ABC in the bipartition A|BC). 
\end{enumerate}
Therefore A and BC can be treated as a pair of qubits in a pure state
and $2\sqrt{\det\rho_{A}}$ taken as the concurrence between A and
BC, obtaining the monogamy inequality \begin{equation}
C_{A(BC)}^{2}\geq C_{AB}^{2}+C_{AC}^{2}.\end{equation}
 This inequality allows one to define the \emph{residual entanglement}
as \begin{equation}
\tau_{ABC}^{c}=C_{A(BC)}^{2}-C_{AB}^{2}-C_{AC}^{2}.\end{equation}
 As shown in their work, the residual entanglement does not depend
on which qubit is chosen as A (the focus) and can thus be considered,
as remarked by the authors, as a \emph{collective property of the
three qubits, that measures an essential three-qubit entanglement}.
The GHZ state has $\tau_{ABC}^{c}=C_{A(BC)}^{2}=1$, while a generalized
W state, $\alpha|001\rangle+\beta|010\rangle+\gamma|100\rangle$,
has $\tau^{c}{}_{ABC}=0$. In this way it is usually said that the
W state only contains bipartite entanglement, while the GHZ state
only contains genuine tripartite entanglement. Note however that both
states can be considered as genuine tripartite entangled as none of
them can be written in an separable way, whatever bipartition is used.

Before proceeding to the case of mixed states it is worth remarking
some points which will be relevant. The concurrence of a pure state
of two qubits is related to the linear entropy, $S_{L}$, of one of
the sub-systems$^{\text{2}}$\footnotetext[2]{Some of the attempts to generalize the concept of concurrence to higher state space dimensions seem to preserve such equivalence \citep{Wootters2001}. I think one could even say that this equivalence can be used to such a generalization. Note that for pure states 
$S_L(\rho_A)$ is a valid measure of entanglement between A and B.}\[
C_{AB}=2\sqrt{\det\rho_{A}}=\sqrt{S_{L}\left(\rho_{A}\right)}=\sqrt{2\left(1-\text{tr}\left[\rho_{A}^{2}\right]\right)}.\]
 Another point that I should mention is that most of the difficulties
in obtaining an entanglement measure for mixed states, comes from
the fact that these are usually defined as convex roof extensions
of a pure state measure. Suppose, for example, that we have a well
defined measure of entanglement for pure states, $E\left(\psi\right)$.
Then the entanglement of a mixed state $\rho$ is defined as \[
E\left(\rho\right)=\inf_{\left\{ p_{i},\phi_{i}\right\} }\sum_{i}p_{i}E\left(\phi_{i}\right),\]
with the infimum taken over all possible decompositions of $\rho$
in a mixture of pure states, $\rho=\sum_{i}p_{i}|\phi_{i}\rangle\langle\phi_{i}|$.
The concurrence of a mixed state of two qubits, for example, is the
convex roof of the pure state concurrence, or equivalently, the convex
roof of the square root of the linear entropy, and in the especial
case of two qubits an analytical expression can be obtained.

That said, let me return to the monogamy inequality, but for mixed
states. At first, when the three qubits are in mixed state $C_{A(BC)}$
is not directly defined as all the four dimensions of BC might be
used. Nonetheless a convex roof extension of the squared concurrence,
$\langle C_{A(BC)}^{2}\rangle^{min}=\inf_{\left\{ p_{i},\phi_{i}\right\} }\sum_{i}p_{i}C_{A(BC)}^{2}\left(\phi_{i}\right)$,
can be used to show that \citep{Coffman2000} \begin{equation}
\langle C_{A(BC)}^{2}\rangle^{min}\geq C_{AB}^{2}+C_{AC}^{2}.\end{equation}
 From this inequality we can also define a residual entanglement for
mixed states of three qubits:

\begin{equation}
\tau_{ABC}^{c}=\langle C_{A(BC)}^{2}\rangle^{min}-C_{AB}^{2}-C_{AC}^{2}.\end{equation}
 However, in general, it will depend on which qubit was chosen as
the focus. Thus for the case of mixed states it is necessary to use
the average over the three possibles focus as the definition of residual
entanglement, and that will be my procedure. From now on I will only
use $C_{A(BC)}^{2}$ as the squared concurrence between A and BC,
and when the state is mixed it is implicit that the convex roof extension
was taken to obtain $C_{A(BC)}^{2}$.

The monogamy inequality can also be established for the negativity.
As show by Ou and Fan in 2007 \citep{Ou2007}, for pure states of
three qubits \begin{equation}
N_{A(BC)}^{2}\geq N_{AB}^{2}+N_{AC}^{2}.\end{equation}
 In the case of mixed states, despite $N_{A(BC)}^{2}$ being well
definied without a convex roof procedure$^{\text{3}}$\footnotetext[3]{In \citet{Lee2003} the convex roof of the negativity is explored as an entanglement measure. This is equal to the concurrence for two qubits, given the equivalence of the concurrence and negativity for pure states with Schmidt rank 2, and can be seen as a generalized concurrence for higher dimensional systems.},
one has to use $\langle N_{A(BC)}^{2}\rangle^{min}$ to obtain,\begin{equation}
\langle N_{A(BC)}^{2}\rangle^{min}\geq N_{AB}^{2}+N_{AC}^{2}.\label{eq: Res Ent Neg}\end{equation}
 Likewise the case of the concurrence Eq. \ref{eq: Res Ent Neg} allows
the definition of a residual entanglement in terms of the negativity:\begin{equation}
\tau_{ABC}^{N}=\langle N_{A(BC)}^{2}\rangle^{min}-N_{AB}^{2}-N_{AC}^{2}.\end{equation}
 Contrary to the case of the concurrence this residual entanglement
depends on which qubit is chosen as $A$, even for pure states. Again,
there is the possibility of taking the average over the three possibles
choices for the focus as the tripartite entanglement measure and,
from now on, that is what I mean by $\tau_{ABC}^{N}$. And it can
be shown that this average is an entanglement monotone \citep{Ou2007}
for pure states. Since for pure states with Schmidt rank 2, which
is our case, the negativity is equivalent to the concurrence, then
$\langle N_{A(BC)}^{2}\rangle^{min}=\langle C_{A(BC)}^{2}\rangle^{min}$
and I refer to {$\langle N_{A(BC)}^{2}\rangle^{min}$} as $C_{A(BC)}^{2}$.
Contrary to the case of concurrence, the residual entanglement in
terms of the negativity, $\tau_{ABC}^{N}$, is not null for the W
state. Latter I will show that $\tau_{ABC}^{N}$ is not a good entanglement
quantifier for mixed states, since it can increase under the action
of local operations.

The difficultty of studying the residual entanglement for mixed states,
those that originate from decoherence, is to obtain an analytical
expression for $C_{A(BC)}^{2}$. Fortunately, in 2002, Osborne obtained
such analytical expression for rank two mixed states \citep{Osborne2005}.
Note that the convex roof of the squared concurrence, in general,
is not equal to the square of the convex roof of the concurrence:
$\langle C^{2}\rangle^{min}\neq\left(\langle C\rangle^{min}\right)^{2}$.
In fact as the pure state squared concurrence is a convex function
then $\langle C^{2}\rangle^{min}\geq\left(\langle C\rangle^{min}\right)^{2}$
(the same is also true for the negativity). In the case of two qubits
the equality was also shown by Osborne. Here I use the expression
obtained by Osborne to study the residual entanglement in terms of
the negativity and concurrence of some states obtained after a decoherence
channel has acted. I do not provide the details of how to obtain this
convex roof neither its final expression, referring the reader to
the original article.

It is important to remark that another possible definition for the
residual entanglement is to use the convex roof extension of the pure
state residual tangle. This is different from the one employed here,
has the advantage of being independent of which qubit is chosen as
A, and was analyzed for a mixture of GHZ and W states with interesting
results in \citep{Lohmayer2006}.

For the sake of completeness let me remark some other points about
the monogamy inequality, before moving to the decoherence models I
use. It is pleasant in the sense that it seems to reveal an entanglement
structure: the entanglement of A with BC can be manifested in the
form of bipartite entanglement with B and C taken individually, plus
an essential three-way entanglement envolving all the three qubits.
Besides this, it can be generalized for a system of $N$ qubits \citep{Osborne2006}.
Nonetheless, the inequality is not obeyed by all entanglement measures,
being a negative example the Entanglement of Formation \citep{Coffman2000},
neither by higher dimensional systems \citep{Ou2007b}. In any case,
there are some attempts to explore this inequality and derivations
from it to establish multipartite entanglement measures for system
of more than three particles \citep{Wong2001,Yu2005,Yu2006,Cai2006,Ou2007a,Ou2007b,Bai2007,Adesso2007},
with some progress and some drawbacks. Nowadays, in my opinion, it
is not totally clear the real power and relevance of these proposals.

\section{Decoherence Models}

It would be desirable to investigate the three paradigmatic types
of decoherence: depolarization, phase damping and amplitude damping.
Of course, I also would like to apply these channels to general three
(actually N) qubits states. But I have to restrict to the cases where
the decohered state is rank 2, since this is the only case where an
analytical expression for $C_{A(BC)}^{2}$ of mixed states is known.
One possibility would be to use mixtures of generalized GHZ and W
states, as they all have rank 2. The problem is that even these simple
states, when suffer the influence of any of these three decoherence
channels locally (each qubit is coupled to an independent channel)
have their rank increased (note that a general mixture of three qubits
can have rank $2^{3}$). Even in the case of a pure GHZ or W state,
the only exception is the local phase damping channel for the generalized
GHZ state and the local amplitude damping for the generalized W state,
which take their rank 1, as they are pure states, and changes it to
2 (see Tab. I).

\begin{table}[ht]
\begin{tabular}{c c c}  & \;\;\; GHZ  & \;\;\;  W 
\tabularnewline \hline \hline
DEP &  \textcolor{red}{>2}        &  \textcolor{red}{>2}    
\tabularnewline
PD  &  \textcolor{green}{\bf{2}}  &  \textcolor{red}{>2} 
\tabularnewline
AD  &  \textcolor{red}{>2}        &  \textcolor{green}{\bf{2}}
\tabularnewline \hline \hline  
\end{tabular} 
\caption{Decoherence Models and the rank of the mixed state generated when these are applied to generalized GHZ and generalized W pure states. DEP stands for Depolarization, PD for phase damping, and AD for amplitude damping. Here every particle is acted localy by a independent channel. Only the two cases where the rank is 2 are studied, as this is the only situation where an analytical expression for $C_{A(BC)}^2$ is know}
\label{tab:Decoherence Models} 
\end{table}

I study the tripartite and bipartite entanglement in the two exceptional
cases mentioned above$^{4}$\footnotetext[4]{Another possibility is to apply the decoherence channel to only one of the qubits, but already in this much simpler situation just the amplitude damping channel when applied to generalized GHZ and W states generates rank 2 mixtures. However this last example will not be shown here, since it does not give qualitative difererences with the presented case.}
using the monogamy relation. Unfortunately, even though the GHZ and
W states are paradigmatic tripartite entangled states, their distribution
of entanglement is somehow trivial: the GHZ does not contain any bipartite
entanglement between the individual parts, so that its tripartite
entanglement is equal to the entanglement between A and BC. In contrast
the W state only contains bipartite entanglement in the sense that
the inequality is saturated and the residual entanglement is null
(using the negativity the inequality is not saturated for the W state).
Because the decoherence channels only act locally and in an independent
way, they are not able to create entanglement between the individual
parts. Thus, the only possibility of having a significant different
behaviour for the tripartite entanglement is to have rather different
decays for the differents bipartitions.

Decoherence and dissipation emerge when we couple the system of interest
with another external system (usually called environment, reservoir,
bath or even ancilla) and only analyse the dynamics of the system
of interest, ignoring the degrees or freedom of the external system.
Mathematically these degrees of freedom are traced out and we end
up with the reduced density matrix of the system of interest. The
dynamics of this open system, which is influenced by the reservoir,
can then be non-unitary and set in decoherence and/or dissipation.

There are many formalisms to treat decoherence and dissipation in
Quantum Mechanics. Here I will use the quantum operator formalism$^{5}$\footnotetext[5]{See chapter 8 of \citep{Nielsen2000} for more details and limitations. Recently results on the generality of completely positive maps as descriptions of dynamical process have been obtained in \citep{Shabani2009}.},
which describe the dynamics experienced by the system as a completely
positive trace preserving (CPTP) linear map that acts on density operators:
$\varepsilon\left(\rho\right)$. This map can include not only unitary
evolutions, but also open ones and general measurements, besides other
general operations. It turns out that any CPTP linear map can be written
in an operator-sum representation: $\varepsilon\left(\rho\right)=\sum_{i}E_{i}\rho E_{i}^{\dagger}$
with $\sum_{i}E_{i}E_{i}^{\dagger}=I$, being this last equality (a
completeness relation) that guarantees the trace preserving property.
The $E_{i}$ are operators on the state space of the system of interest,
usually called as Kraus operators, and can be obtained from the knowledge
of the Hamiltonian governing the whole system (interest + environment).

The amplitude channel can represent a typical interaction of a qubit
with a zero-temperature reservoir in its ground state $|0\rangle_{R}$.
In this interaction there is a finite probability, $p$, that the
upper state $|1\rangle$ of the qubit decays to $|0\rangle$ and creates
one excitation in the reservoir, which finishes at $|1\rangle_{R}$.
Of course there is a probability $1-p$ that nothing happens. This
interaction can represented by the map \begin{eqnarray*}
|0\rangle|0\rangle_{R} & \to & |0\rangle|0\rangle_{R}\\
|1\rangle|0\rangle_{R} & \to & \sqrt{1-p}|1\rangle|0\rangle_{R}+\sqrt{p}|0\rangle|1\rangle_{R},\end{eqnarray*}
 and has the followings Kraus operators: $E_{1}=\left(|0\rangle\langle0|+\sqrt{1-p}|1\rangle\langle1|\right)$
and $E_{2}=\sqrt{p}|0\rangle\langle1|$. The probability $p$ is related
to the interaction time and under Markovian approximation $p=1-e^{-\Gamma t}$
with $\Gamma$ a constant which depends on characteristics of the
environment and its coupling with the system.

The phase damping channel could also represent an interaction of the
qubit with a reservoir. But, contrary to the amplitude damping, there
is no loss of energy, but only information loss (in this sense the
process is uniquely quantum). The eigenstates of the qubit do not
change, but accumulate an unknown phase that destroys the relative
phase between them (loss of information). It represents, for example,
the elastic scattering between the qubit and the reservoir and is
described by the following map \begin{eqnarray*}
|0\rangle|0\rangle_{R} & \to & |0\rangle|0\rangle_{R}\\
|1\rangle|0\rangle_{R} & \to & \sqrt{1-p}|1\rangle|0\rangle_{R}+\sqrt{p}|1\rangle|1\rangle_{R}\end{eqnarray*}
 Note that the eigenstates of the qubit are not changed, but become
entangled with the reservoir, and this will cause the loss of coherence
in the qubit when the reservoir is traced out. The Kraus operators
for the phase damping are: $E_{1}=\sqrt{1-p}\left(|0\rangle\langle0|+|1\rangle\langle1|\right)$,
$E_{2}=\sqrt{p}\left(|0\rangle\langle0|\right)$ and $E_{3}=\sqrt{p}\left(|1\rangle\langle1|\right)$.
In the situation where the channel acts independently on each qubit
of the system the formalism extends directly.

\section{Results}

First I present the behaviour of a generalized GHZ state under the
action of three independent phase damping channels on each qubit.
In Fig. 1 the behavior of the entanglement of A with BC given in terms
of the negativity $N_{A(BC)}$ (dashed curve) and of the convex roof
of the squared negativity $C_{A(BC)}^{2}$ (solid curve) can be seen
as a function of $p$: the convex roof decays faster than the negativity.
However it may be more fair to compare $C_{A(BC)}^{2}$ with the squared
negativity $N_{A(BC)}^{2}$. This is not shown here, as the second
becomes equal to the first (see footnote 3). Note that here the bipartite
entanglement given by the convex roof is equal to the tripartite entanglement
given by the residual entanglement: $\tau_{ABC}^{c}=\tau_{ABC}^{N}=C_{A(BC)}^{2}$.
From the graphics on the right of Fig.1 it can also be checked that
any generalized GHZ has less residual entanglement than the GHZ itself
for any value of $p$ and that the residual entanglement always decrease
with $p$

\begin{figure}
\noindent \begin{centering}
\includegraphics[clip,scale=0.35]{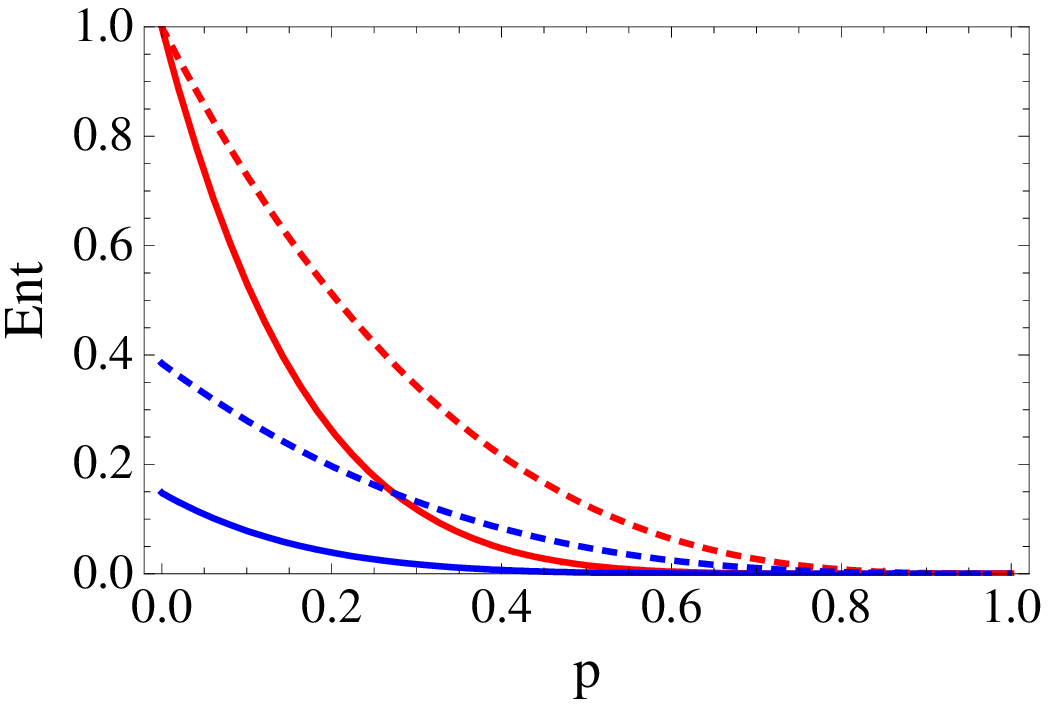}~~\includegraphics[clip,scale=0.35]{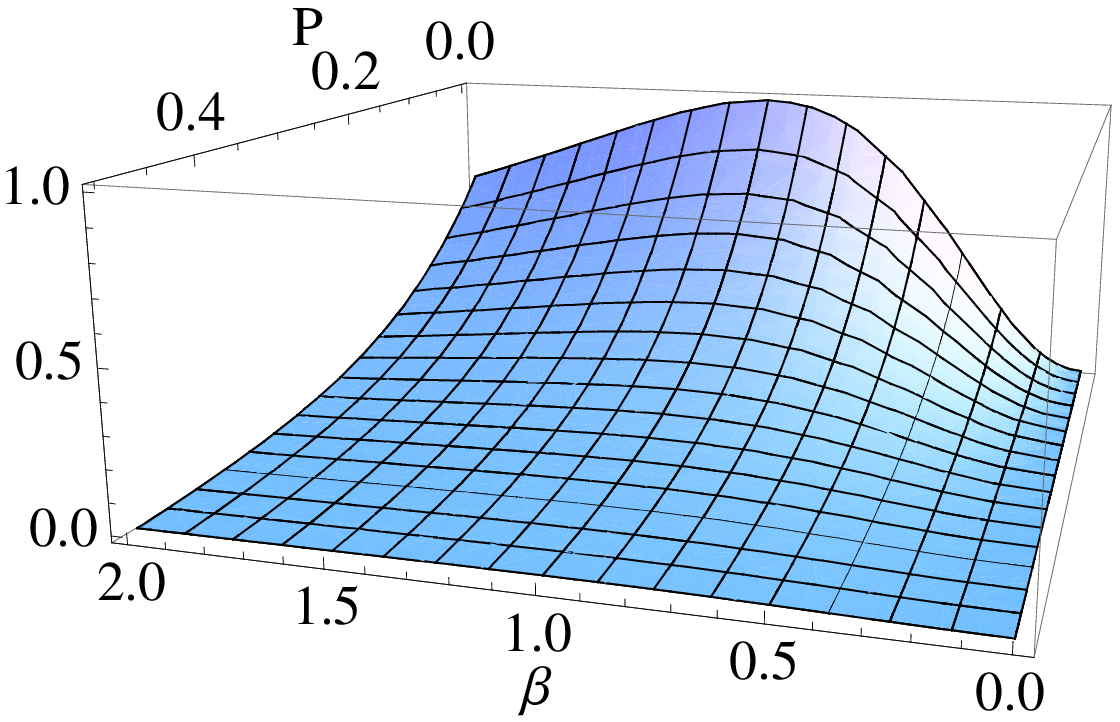}
\par\end{centering}

\caption{(color online) On the left $N_{A(BC)}$ (dashed) and $C_{A(BC)}^{2}$
(solid) for a state obtained when a phase damping channel acts locally
and independently on each of the three qubits of an initial GHZ state
(upper curves) and a generalized one, which without normalization
is $0.2|000\rangle+|111\rangle$. On the right $C_{A(BC)}^{2}$ for
a family of generalized GHZ states, $1|000\rangle+\beta|111\rangle$.
For the two states used on the left and the ones on the right $C_{A(BC)}^{2}=N_{A(BC)}^{2}$.}

\end{figure}

The second possibility is to look at the action of the amplitude damping
in a generalized W state. In this case $N_{A(BC)}^{2}\neq C_{A(BC)}^{2}$,
but they have a qualitative similar behaviour. As an illustration,
Fig. 2 shows both of them for the W state (upper curves in red) and
a generalized one (lower curves in blue), which without normalization
is $|001\rangle+5|010\rangle+10|100\rangle$. Note that the difference
between the two measures increases with decoherence at the beginning
and then goes to zero again. In fact if $N_{A(BC)}^{2}$ is equal
to zero, then $C_{A(BC)}^{2}$ should also be, since there are no
bound entangled states of rank 2 \citep{Pawel2003}.

\begin{figure}
\noindent \begin{centering}
\includegraphics[clip,scale=0.5]{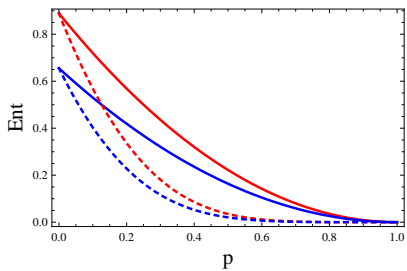}
\par\end{centering}

\caption{(color online) $C_{A(BC)}^{2}$ (solid) and $N_{A(BC)}^{2}$ (dashed)
obtained when an amplitude damping channel acts locally and independently
on each qubit of an initial W state (upper curves) and a generalized
one, which without normalization is $|001\rangle+5|010\rangle+10|100\rangle$.}

\end{figure}

After viewing the difference between $C_{A(BC)}^{2}$ and $N_{A(BC)}^{2}$
I investigate the distribution of the bipartite entanglement, using
the concurrence as the measure of entanglement between qubits. In
Fig. 3 I show this distribution in terms of the concurrence for the
W state (left) and a generalized one (right), which without normalization
is $|001\rangle+2|010\rangle+3|100\rangle$. There the behavior of
the entanglement between one qubit and the rest for differents focus,
$C_{A(BC)}^{2}$, for example, under decoherence can be seen as solid
curves. The entanglement between individual qubits is shown in dotted
curves, while the residual entanglement is not show as it is null
at the begginig and for any value of $p$. This indicates that the
sum of the bipartite entanglement between the qubits decays at the
same rate of the entanglement between one qubit and the rest. Fig.
4 is analogous but in terms of the negativity: we have $N_{AB}^{2}$
instead of $C_{AB}^{2}$. In this case the residual entanglement $\tau_{ABC}^{N}$
(dashed curve) may increase under the action of the local decoherence,
indicating that this cannot be a good entanglement quantifier.

\begin{figure}
\begin{centering}
\includegraphics[clip,scale=0.35]{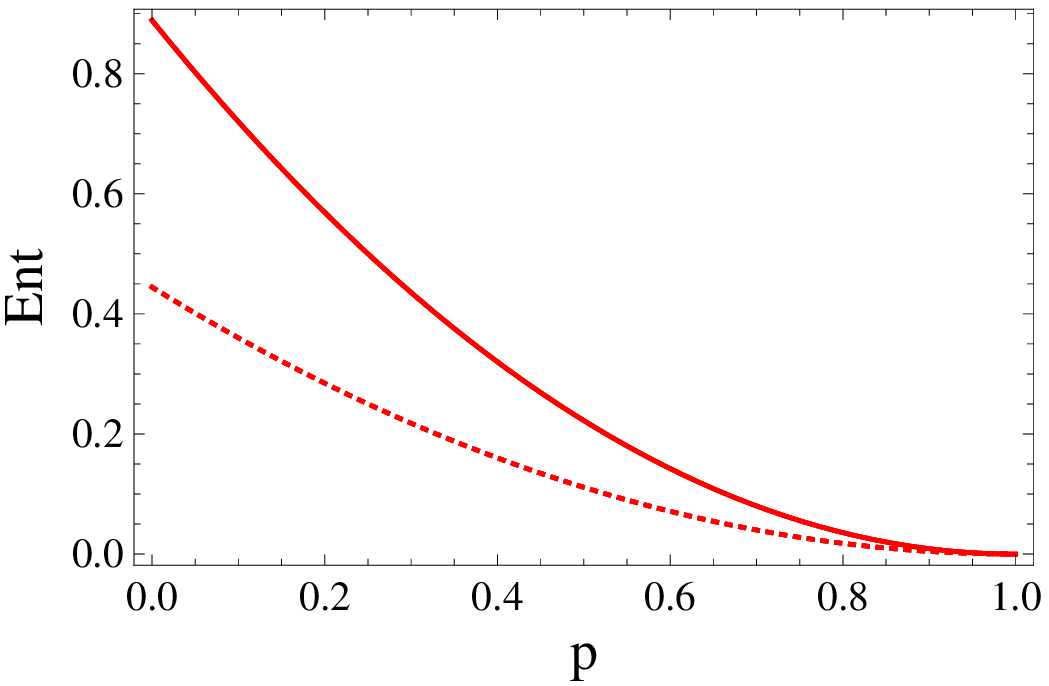}~~\includegraphics[clip,scale=0.35]{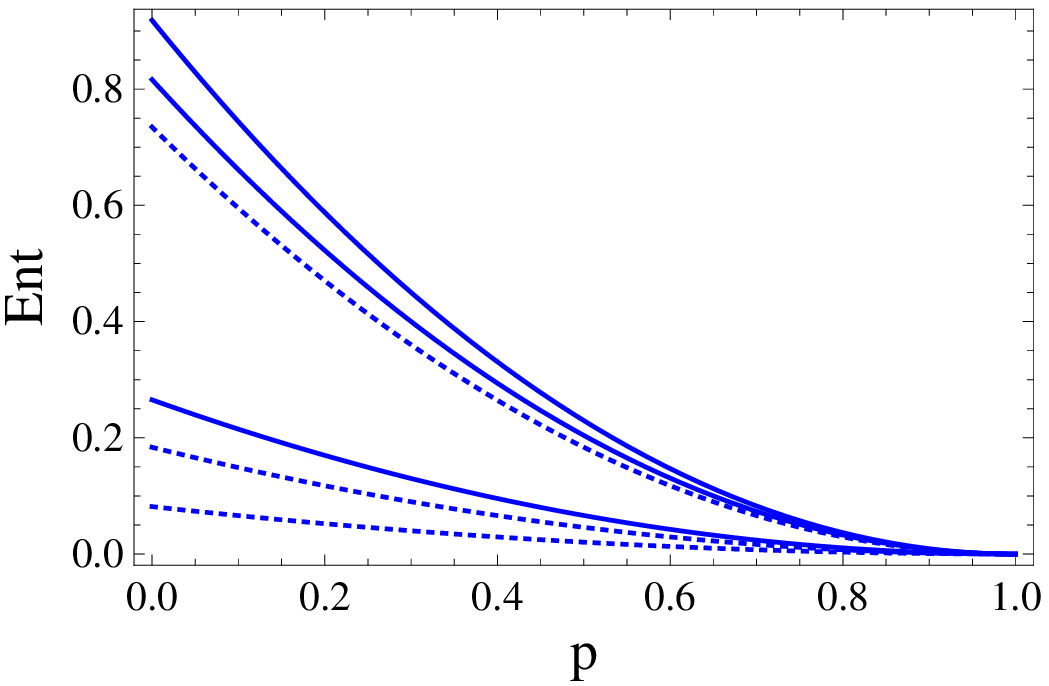}
\par\end{centering}

\caption{(color online) The distribution of entanglement in terms of the concurrence
as a function of $p$, when an amplitude damping channel acts locally
and independently on each qubit of an initial W state (left) and a
generalized one (right), which without normalization is $|001\rangle+2|010\rangle+3|100\rangle$:
$C_{A(BC)}^{2}$, $C_{B(AC)}^{2}$ and $C_{C(AB)}^{2}$ are ploted
as solid lines from top to the bottom, respectively. $C_{AB}^{2}$,
$C_{AC}^{2}$ and $C_{BC}^{2}$ are given by dotted lines from top
to the bottom, respectively. The residual entanglement $\tau_{ABC}^{c}$,
is null for any value of $p$.}

\end{figure}

\begin{figure}
\begin{centering}
\includegraphics[clip,scale=0.35]{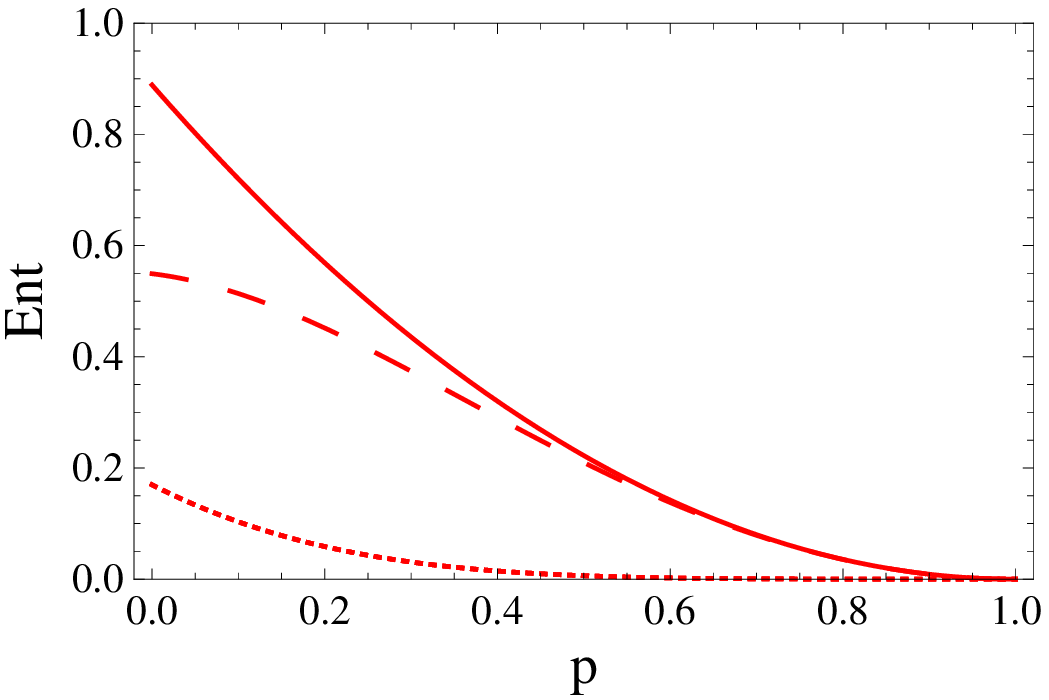}~~\includegraphics[clip,scale=0.35]{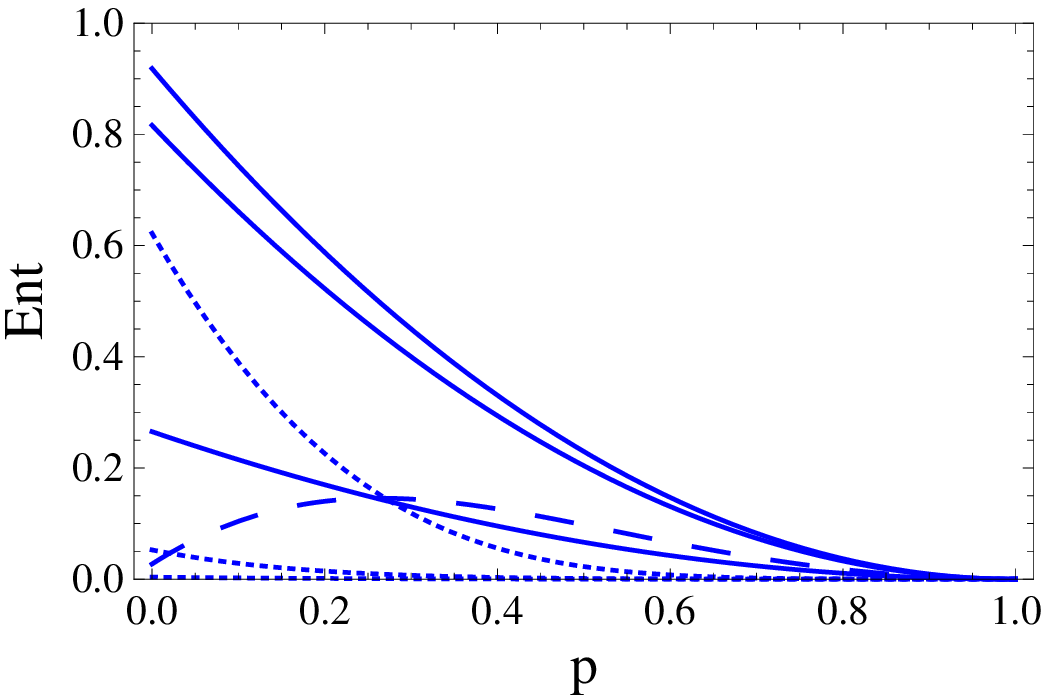}
\par\end{centering}

\caption{(color online) The distribution of entanglement in terms of the negativity
as a function of $p$, when an amplitude damping channel acts locally
and independently on each qubit of an initial W state (left) and a
generalized one (right), which without normalization is $|001\rangle+2|010\rangle+3|100\rangle$:
$C_{A(BC)}^{2}$, $C_{B(AC)}^{2}$ and $C_{C(AB)}^{2}$ are ploted
as solid lines from top to the bottom, respectively. $N_{AB}^{2}$,
$N_{AC}^{2}$ and $N_{BC}^{2}$ are given by dotted lines from top
to the bottom, respectively. The residual entanglement $\tau_{ABC}^{N}$
is show in dashed. Note that it may increase, showing that it cannot
be an entanglement monotone.}

\end{figure}

\section{Conclusions}

In sum, I first investigated the behavior of the squared negativity
and its pure state convex roof extension under some models of decoherence
for the generalized GHZ and generalized W states. While they are equal
for the generalized GHZ state, that is not true for the generalized
W state. However in this last case they exhibit similar qualitative
behavior decaying with the decoherence even tough their difference
can increase.

I also studied how the distribution of entanglement behaves under
these models of decoherence making use of the monogamy inequality
for mixed states in terms of the concurrence and the negativity. For
this aim I compared the behavior of the entanglement between one qubit
and the other two treated as a single system with the one between
individuals qubits. For the generalized W state I found out cases
where the residual entanglement in terms of the negativity can increase
under local decoherence even when all the bipartite entanglement is
decaying, showing that it can not be a good entanglement measure.

In relation to the behaviour of multipartite entanglement under decoherence
these results show no qualitative difference between the bipartite
and multipartite entanglement for generalized GHZ and W states. However
it would be interesting to be able to study these monogamy relations
for states with richer entanglement structure than the GHZ and the
W states. Would it be possible to happen that the bipartite entanglement
between individuals qubits goes to zero while the multipartite one
is still finite, and maybe only goes to zero asymptotically (sudden
death of only bipartite entanglement)? We expect the rate of decay
of both parts to be equal, since the tripartite entanglement should
not increase under local operations. How about non-local environments?

These studies could also increase our understandig of multipartite
entanglement per si. For know, we showed that the residual entanglement
in terms of the negativity as used here can not be considered as a
good entanglement measure. But, as mentioned before, one ca also define
the residual entanglement for mixed states as the convex roof of the
pure state residual entanglement \citep{Lohmayer2006}. And there
are many other proposals of generalizations for higher Hilbert space
dimensions and more than three particles ( \citep{Wong2001,Yu2005,Yu2006,Cai2006,Ou2007a,Ou2007b,Bai2007,Adesso2007},
for example). Therefore, it would be interesting to investigate the
behavior of these others measures under decoherence.

\begin{acknowledgments}
I would like to thank Marcio Cornelio, Gustavno Rigolin and Marcos
César de Oliveira for many interesting and frutiful conversations
on entanglement and monogamy inequality. I am also grateful to Tobias
Osborne for an exchange of ideas about the convex roof of the squared
pure state concurrence and to Gustavo Rigolin for a careful reading
of the manuscript and many suggestions. This work was supported by
Fundação de Amparo a Pesquisa do Estado de São PAulo (FAPESP) under
the project 2008-03557-8.

\bibliographystyle{apsrev}
\clearpage\addcontentsline{toc}{chapter}{\bibname}\bibliography{/home/tro/Refs/References-Formatted}

\end{acknowledgments}

\end{document}